\documentclass{aa}  

\usepackage{graphicx}
\usepackage{txfonts}

\usepackage{subcaption}
\usepackage{lscape}
\usepackage{placeins}

\usepackage{hyperref}
\usepackage{amsmath,amsfonts,amssymb}
\usepackage{multirow}
\usepackage{aasmacros}
\usepackage{xcolor}

\begin{document}

\title{Temporal regularization of petal modes in single conjugate adaptive optics systems}
\subtitle{The soft-limiter approach}

\author{Guido Agapito\inst{1,4}\corrauth{guido.agapito@inaf.it}
        \and Jueqi Lin\inst{2}
        \and Carmelo Arcidiacono\inst{3,4}
        \and Giulia Carlà\inst{1,4}
        \and Enrico Pinna\inst{1,4}
        \and Fabio Rossi\inst{1,4}}

\institute{INAF Osservatorio Astrofisico di Arcetri, Largo E. Fermi 5, 50125 Firenze, Italy
           \and
           School of Automation, Southeast University, 2 Sipailou, Nanjing 210096, China
           \and
           INAF - Osservatorio Astronomico di Padova, Vicolo Osservatorio 5, 35122 Padova, Italy
           \and
           ADaptive Optics National laboratory in Italy (ADONI)}

\date{Received XX May 2026; accepted XX XX 2026}

\abstract
  {The next generation of extremely large telescopes (ELTs) relies on segmented mirrors and thick support structures that fragment the optical pupil of the instrument.}
  {In single conjugate adaptive optics (SCAO) systems, this fragmentation leads to the emergence of differential piston errors, or petal modes. The pyramid wavefront sensor (PWFS), baseline for instruments such as ANDES, MICADO, HARMONI, and GMT-NGSAO, suffers from poor sensitivity to these petal modes when operating at visible wavelengths under realistic residual turbulence. Our aim is to prevent the resulting ``island effect,'' where poorly observed differential pistons accumulate and degrade the point spread function, without compromising the correction of continuous atmospheric modes.}  
  {We propose a deterministic and computationally efficient control strategy: the minimum mean square error (MMSE) soft-limiter. Rather than relying on an instantaneous spatial reconstruction of the phase steps, we utilized a subspace MMSE estimator derived from von Kármán statistical priors coupled with a continuous temporal regularization. This acts as a localized leaky integrator applied strictly to the relative petal degrees of freedom.}
  {End-to-end SPECULA simulations for the ELT ANDES instrument show that while a classical integrator loses lock and wanders across multiple fringes under degrading seeing ($\ge 1.0''$), the MMSE soft-limiter confines the differential piston within a single fringe across the full $0.6-1.2''$ range. This prevents control-induced divergence, bounds the petal excursion, and reduces its science-band impact under severe turbulence while retaining optimal continuous-mode correction at good seeing without the steady-state Strehl penalty incurred by spatial mitigation strategies such as edge actuator coupling. We theoretically derived and numerically verified the operating window for the leakage gain, $g \lesssim g_{\mathrm{eff}}/\pi$. Below this threshold, the fast loop retains the full fractional range of genuine physical steps, up to $\lambda_{WFS}/2$ with a small droop; above it, retention degrades gradually.} 
  {This temporal regularization maintains a stable modulo-$\lambda_{WFS}$ wavefront on the fast PWFS, allowing for a slower phasing sensor to resolve integer ambiguities. Because the operating window narrows with turbulence, the boundaries cross at roughly $1.0''$ seeing. Beyond this point, retaining physical discontinuities becomes the exclusive responsibility of the phasing channel---a key design driver for segmented ELT-class facilities.}


\maketitle
\nolinenumbers

\section{Introduction}
\label{sec:intro}
The design of forthcoming facilities, such as the European Extremely Large Telescope (ELT; \citealt{2024Msngr.192....3C}), the Giant Magellan Telescope (GMT; \citealt{2024SPIE13094E..17B}), and the Thirty Meter Telescope (TMT; \citealt{2025AAS...24514507L}), introduces the challenge of segmented pupils to adaptive optics (AO) control \citep{Schwartz2018, BertrouCantou2020, Wang2024}. Mechanical structures such as spider arms, often wider than the atmospheric coherence length $r_0$, mechanically divide the pupil into disconnected fragments, enabling the emergence of differential piston errors \citep{2022A&A...665A.158P}.

For single conjugate AO (SCAO) instruments such as ANDES \citep{Selmi2025}, HARMONI \citep{Bond2022}, MICADO \citep{2022SPIE12185E..4SC}, and GMT-NGSAO \citep{Pinna2014}, the modulated pyramid wavefront sensor (PWFS; \citealt{1996JMOp...43..289R}) is the primary sensor. The control of PWFS-based systems has seen significant progress over the past two decades. Early theoretical and laboratory works demonstrated its inherent capability for segmented mirror alignment~\citep{2005OptL...30.2572E, 2008SPIE.7012E..3DP}. On-sky, advanced control strategies have been developed to master the sensor nonlinearities and optical gain variations, enabling the precise compensation of noncommon path aberrations (NCPA) \citep{2020A&A...636A..88E} and the implementation of robust temporal filtering, as successfully demonstrated by the SOUL upgrade at the Large Binocular Telescope \citep{2021arXiv210107091P, 2021MNRAS.508.1745A}. Furthermore, recent studies exploring the nonmodulated PWFS regime have highlighted its superior sensitivity to differential piston modes \citep{2023A&A...677A.168A}, making it a highly attractive option for phasing fragmented pupils.

However, despite these sophisticated control frameworks, the PWFS faces severe challenges when dealing with fragmented apertures under realistic observing conditions. In the partial correction regime, the sensor suffers from a dynamic drop in sensitivity---the well-known optical gain effect---which depends heavily on the instantaneous level of residual atmospheric turbulence \citep{2020A&A...636A..88E, 2023A&A...677A.168A}. While different studies report varying degrees of severity for this effect, they all agree that the petal sensitivity $\eta_p$ systematically degrades with increasing residual phase variance; the quantitative discrepancies between specific numerical evaluations (e.g., \citealt{2023A&A...677A.168A} versus \citealt{BertrouCantou2022}) primarily reflect the differences in the simulated modulation radius and spider width. Furthermore, the PWFS response to phase steps across gaps is inherently nonlinear and exhibits a sinusoidal behavior \citep{2002ESOC...58..161E}. This combination of sensitivity degradation, nonlinear saturation, and severe modal cross-talk makes the stable closed-loop control highly challenging. Consequently, the fast SCAO loop is prone to instabilities, triggering the ``island effect'' where the deformable mirror (DM) accumulates unphysical pistons that significantly degrade the point spread function \citep{Schwartz2018}.

To mitigate this, various solutions have been proposed. On the control software side, strategies range from edge actuator coupling (EAC) for HARMONI \citep{Schwartz2018, Bond2022} to continuous modal bases \citep{BertrouCantou2020}. On the hardware side, modifications to the fast loop, such as spatially filtered unmodulated pyramids, have been investigated \citep{Levraud2024}. 

Furthermore, to completely resolve the absolute $2\pi$ phase ambiguity and measure physical steps (e.g., from the low-wind effect; \citealt{2018SPIE10703E..2AM}), the community has increasingly turned to secondary focal-plane sensors. Pioneering focal-plane phase-diversity algorithms such as ``Fast \& Furious'' were initially developed to tackle spider-induced discontinuities on monolithic 8-m class telescopes such as VLT/SPHERE and Subaru/SCExAO \citep{Wilby2016_FastFurious,2020A&A...639A..52B}. Building on these concepts for the 30-m class segmented telescopes, focal-plane fringe tracking and interferometric techniques operating at low frequencies ($\sim 1\,\mathrm{Hz}$) have emerged as the standard. The holographic dispersed fringe sensor (HDFS) has been successfully validated for the GMT \citep{Haffert_HDFS, QuirosPacheco2022} and proposed for the TMT to mitigate petalling in high-contrast imaging instruments \citep{Wang2024}, complementing interferometric concepts demonstrated at Subaru/SCExAO \citep{Vievard_SCExAO}. Within the ELT framework, significant efforts are dedicated to focal-plane sensing techniques such as the Broyden–Fletcher–Goldfarb–Shanno (BFGS) algorithm \citep{2017JATIS...3c9001L} or the linearized focal-plane technique (LIFT)---whose application has been extensively studied for both GMT \citep{Cheffot2022} and ELT instruments such as MORFEO \citep{Plantet2026}--- or machine learning based approaches \citep{2022SPIE12185E..5DR,2025A&A...704A.246J} as well as novel concepts such as the CiaoCiao wavefront sensor (WFS) \citep{CiaoCiao}.

While these dedicated phasing sensors are vital, they require the fast SCAO loop to remain stable and not inject spurious control-induced pistons during their integration time. In this paper, we present an alternative, software-based approach for the fast loop: a temporal regularization technique called the minimum mean square error (MMSE) soft-limiter. Conceptually rooted in well-established control theory and signal processing paradigms---sharing mathematical similarities with selective anti-windup mechanisms and Tikhonov ($L_2$) regularization---this method constrains the divergence of the petal modes by exploiting the temporal dynamics of the closed loop. Instead of modifying the WFS optical path or altering the DM influence functions, we introduce a localized forgetting factor strictly applied to the poorly observed differential piston states, effectively bridging the gap between fast atmospheric correction and slow interferometric phasing.

\section{The differential piston challenge: Definition, modal correlation, and estimation complexity}
\label{sec:challenge}

In AO systems employing segmented DMs, the introduction of phase discontinuities between segments represents a critical challenge for wavefront control \citep{2000ExA....10....5G, 2014MNRAS.440..125P}. To avoid the terminological ambiguities common in the literature, we explicitly define our naming convention, following \citet{BertrouCantou2022}. We use the term ``segment piston'' to denote the spatially uniform phase over a single pupil fragment (a petal). Because a global uniform phase is unobservable, the absolute segment piston has no physical meaning. What the system is sensitive to is the differential piston, defined as the pure phase step between a segment, $i$ and a chosen reference segment, $\Delta P_i = a_i - a_{\text{ref}}$, completely devoid of continuous deformations. In this work, the terms ``differential piston'' and ``petal mode'' \citep{Levraud2024} are used interchangeably to describe these pure phase steps.

\begin{figure}[h]
    \centering
    \resizebox{\hsize}{!}{\includegraphics{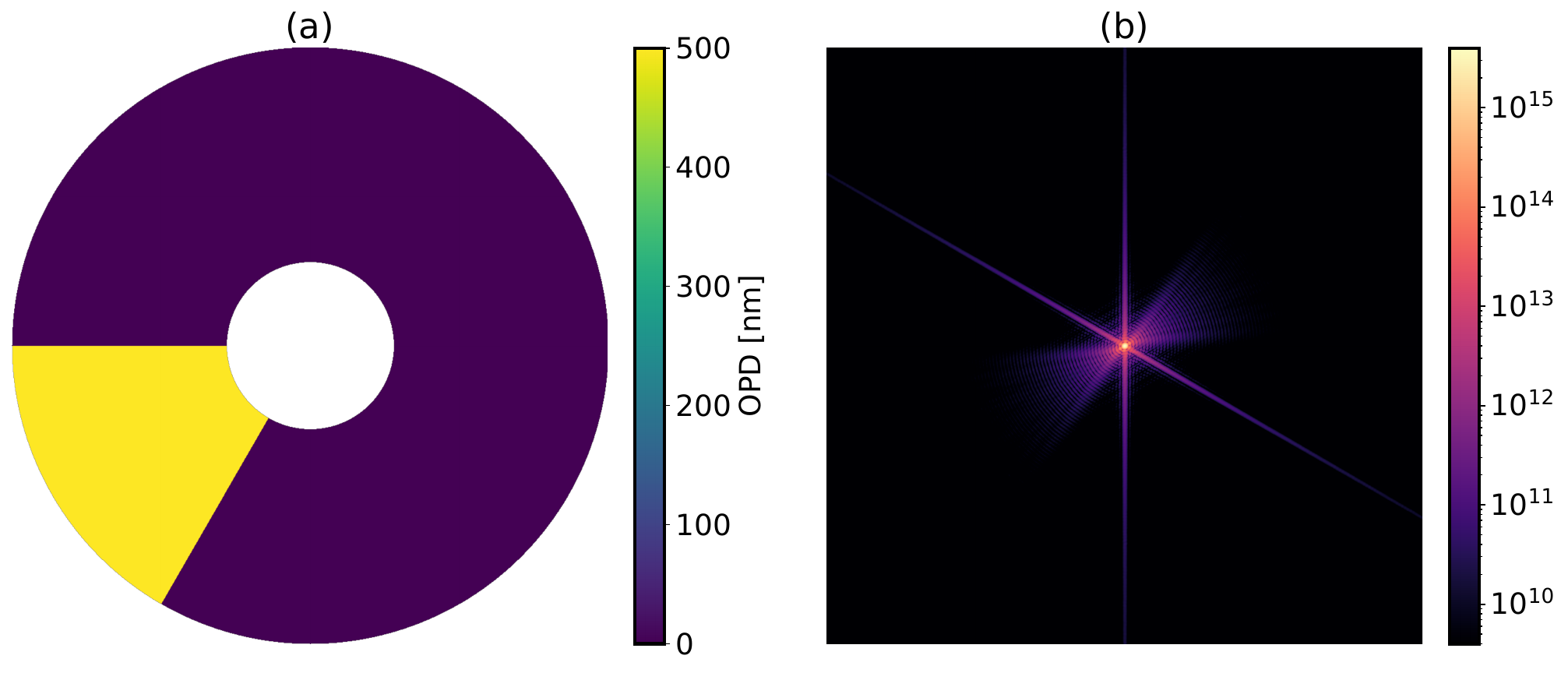}}
    \caption{Pure petal mode in the spatial and frequency domains. \textbf{(a)} Spatial representation of a pure petal mode. \textbf{(b)} Corresponding power spectrum, computed via fast Fourier transform, highlighting the broadband nature of the discontinuity.}
    \label{fig:petal_fft}
\end{figure}

From a frequency domain perspective, a step discontinuity generates a broad spectrum that includes spatial components well above the Nyquist frequency dictated by the DM actuator grid (see Fig.~\ref{fig:petal_fft}).

This broadband nature makes the differential piston inherently nonorthogonal to standard continuous modal bases, such as Zernike polynomials or Karhunen-Loève modes, as the petal pattern has a nonzero projection onto the continuous basis. We stress that this is a deterministic, geometric property of the modal patterns, and not a statistical correlation between the underlying random amplitudes---a distinction that becomes important for the estimator introduced in Sect.~\ref{sec:implementation}. Because of this nonorthogonality, separating the pure differential piston from continuous atmospheric turbulence is not trivial. Continuous low-order modes, such as tilt or coma, naturally introduce a mean height difference across a given segment, which we refer to as the mean segment phase.

\begin{figure}[h]
    \centering
    \resizebox{\hsize}{!}{\includegraphics{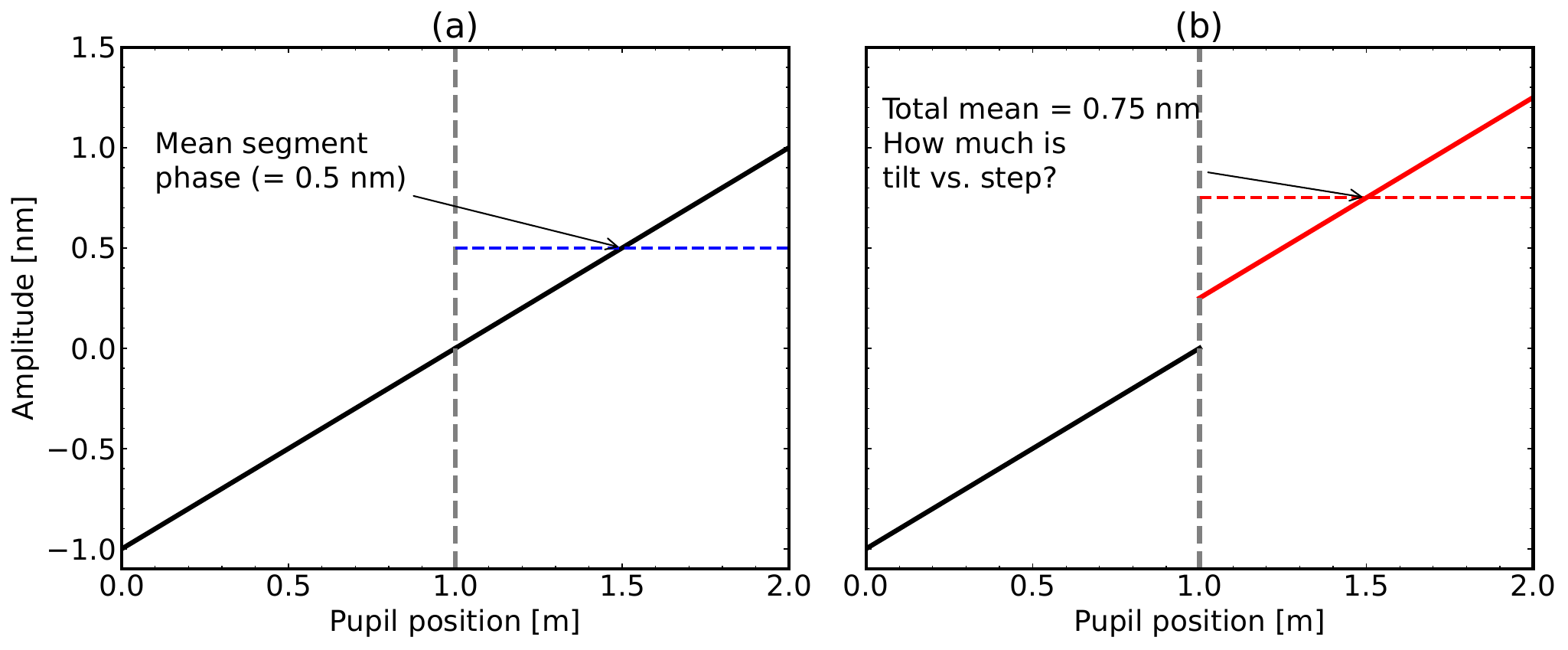}}
    \caption{Modal correlation problem. One-dimensional pupil cross-section demonstrating the ambiguity between continuous and discontinuous aberrations. \textbf{(a)} Continuous tilt generates a mean height difference over the right segment, defined as mean segment phase. \textbf{(b)} Pure phase step (segment piston) superimposed on the tilt.}
    \label{fig:correlation}
\end{figure}

By definition, a mean segment phase induced by a continuous mode and a pure phase step represent the same subspace projection when averaged over the segments (see Fig.~\ref{fig:correlation}). While the mathematical separation between the two would be perfectly invertible if the projection basis contained both the exact discontinuities and the full continuous wavefront space, in practical control systems we operate on truncated bases. In this regime, the separation becomes an ill-conditioned inverse problem; projecting a continuous phase containing tilt onto a truncated petal basis erroneously interprets the mean segment phase as a pure phase step.

This mathematical ambiguity reflects directly and severely on the design and evaluation of dedicated WFSs. The major difficulty lies in defining the ``ground truth.'' Due to the inability to trivially isolate the discontinuity component, it becomes extremely complex to establish whether the sensor is actually measuring the desired parameter \citep{BertrouCantou2022}. To resolve this ambiguity and evaluate sensor performance, it is necessary to impose a priori constraints, such as using MMSE estimators based on temporal statistics, spatial derivative-based filters that ignore edge zones, or machine learning architectures trained to recognize the specific signature of the step \citep{rossi2026b}.

The problem of isolating differential piston modes is formally analogous to edge-preserving denoising in signal processing. During the preliminary phase of our study, we explored nonlinear and edge-aware spatial estimators designed to exploit the sparse, discontinuous nature of petal modes. For instance, total variation (TV) regularization \citep{Rudin1992} and data-driven approaches such as convolutional neural networks (CNNs; e.g., U-Net, \citealt{Ronneberger2015}) outperform linear baselines in single-frame absolute petal extraction accuracy \citep{rossi2026b}.

However, these static gains carry an unsustainable practical cost for the fast loop. Genuine TV regularization requires iterative minimization schemes \citep{Chambolle2011, 10.1561/2200000016, Gfrerer2026} that reach convergence on timescales many orders of magnitude beyond the strict $\sim 1\,\mathrm{ms}$ per-frame budget of modern SCAO real-time computers (RTC), such as the HEART framework \citep{Smith2022_HEART}. Similarly, CNNs rely on heavier inference engines.

By contrast, the linear MMSE estimator is attractive precisely for its simplicity and minimal computational footprint since it reduces to a single, highly optimized matrix-vector multiplication (MVM) that comfortably fits submillisecond latency constraints. While a linear estimator inherently suffers from spatial limitations---most notably the Gibbs ringing that any truncated continuous basis exhibits at a sharp step---our key observation is that the closed-loop instability stems from the uncontrolled temporal accumulation of poorly observed differential pistons, not from the ringing of any individual reconstruction.

Rather than attempting to improve the instantaneous spatial estimate, we therefore leave the computationally efficient per-frame MMSE untouched and shift the regularization paradigm from the spatial to the temporal domain, constraining how the petal modes are allowed to evolve from frame to frame. In the remainder of the paper, we develop this temporal regularization and show that it stabilizes the loop without degrading the continuous atmospheric correction.

\section{SCAO architecture and the MMSE soft-limiter}
\label{sec:implementation}

The ANDES SCAO system is designed around a dual-loop paradigm:
\begin{enumerate}
    \item The fast loop: A modulated PWFS operating at up to $1\,\mathrm{kHz}$ corrects continuous atmospheric turbulence using $\sim 4000$ continuous modes. The standard temporal control for this modal basis is a simple integrator, which greatly simplifies the overall computational framework. The only exception is the tip-tilt control, which employs a dedicated infinite impulse response (IIR) filter to effectively reject the significant wind-shake introduced by the ELT M2 mirror, whose root mean square (RMS) amplitude reaches $11\,\mu\mathrm{m}$.
    \item The slow loop (phasing channel): The slow loop is the dedicated sensor operating at $\sim 1\,\mathrm{Hz}$ in the near-infrared, employing focal-plane phase retrieval or interferometric techniques (such as LIFT at two wavelengths) to resolve the absolute $2\pi$ phase ambiguity and measure the true physical steps (e.g., low-wind effect). This slow loop is not modeled in this work.
\end{enumerate}

The slow phasing channel efficiently rejects the segment piston up to a bandwidth of a fraction of a Hz. It is thus necessary that the fast loop does not inject spurious segment piston variations above this frequency.

In standard SCAO control, the continuous atmospheric modes are governed by a classical integrator, which computes a provisional modal command state $\mathbf{m}_{k+1}^*$ at step $k+1$ as
\begin{equation}
    \label{eq:integrator}
    \mathbf{m}_{k+1}^* = \mathbf{m}_{k} + g_{c} \Delta \mathbf{m}_{k},
\end{equation}
where $g_{c}$ is the loop gain and $\Delta \mathbf{m}_{k}$ is the instantaneous modal correction computed from the PWFS residual error.

To prevent the poorly observed differential pistons from diverging within this continuous framework, we implemented the MMSE estimator as a ``soft-limiter'' operating on the accumulated provisional state. We regarded this accumulated command as the superposition of the poorly observed differential-piston content and the continuous atmospheric content,
\begin{equation}
    \label{eq:signal_model}
    \mathbf{m}_{k+1}^* = P\,\mathbf{p}_{k+1} + \boldsymbol{\phi}_{k+1},
\end{equation}
where $\boldsymbol{\phi}$ collects the continuous modes and $P$ is the geometric operator mapping the five differential-piston degrees of freedom into the $\sim 4000$-mode command space. We modeled the random amplitudes $\mathbf{p}$ and $\boldsymbol{\phi}$ as statistically uncorrelated, reflecting their distinct physical origins. Under this model, the linear minimum-mean-square-error estimate of the petal vector is given by the Wiener filter, $\hat{\mathbf{p}}_{k+1} = C_p P^T \left[ P C_p P^T + C_w \right]^{-1} \mathbf{m}^*_{k+1}$, where $C_p$ is the prior covariance of the differential piston amplitudes, and $C_w$ is the continuous atmospheric covariance derived from von Kármán statistics \citep{ishimaru1978volume2}.

To minimize the real-time computational footprint, we defined the leakage matrix $L = g\,P C_p P^T \left[ P C_p P^T + C_w \right]^{-1}$, where $g$ is a fractional leakage gain ($0 < g \ll 1$). Rather than tuning $C_p$ to the expected physical amplitude of the petal modes, we deliberately employed a highly uninformative prior ($\sigma_\mathrm{prior} \to \infty$). In this regime, the specific matrix structure of $C_p$ becomes irrelevant, and the estimator naturally converges to the Gauss-Markov (minimum-variance unbiased) limit. This is a fundamental design choice. By operating in the unbiased limit ($\hat{\mathbf{p}} \approx \mathbf{p}$), the spatial extraction is decoupled from the actual petal amplitude. Consequently, the petal amplitude is not a free parameter of the method, ensuring that $g$ remains the sole tuning parameter of the temporal loop. This makes the static extraction mathematically immune to the optical capture range or phase wrapping---though, as quantified in Sect.~\ref{sec:lwe_retention}, the ability of the closed-loop system as a whole to retain true physical steps does physically depend on the optical sensitivity $\eta_p$.
Since $g$ is small and the associated temporal dynamics are slow, the regularization term can be computed using the command state from the previous iteration without performance loss. 
We defined the leak vector $\mathbf{l}_k = L\,\mathbf{m}^*_k = g\,P\,\hat{\mathbf{p}}_k$, that is, the petal content identified by the estimator, mapped back to the command space and scaled by $g$. In practice it is evaluated as two MVMs of size $5 \times 4000$: the estimate $\hat{\mathbf{p}}_k$, followed by its projection $P\,\hat{\mathbf{p}}_k$. The fast loop command update applied to the DM then becomes
\begin{equation}
    \label{eq:soft_limiter_update}
    \mathbf{m}_{k+1} = \mathbf{m}_{k+1}^* - \mathbf{l}_k = \mathbf{m}_k + g_c \Delta \mathbf{m}_k - \mathbf{l}_k.
\end{equation}
This formulation allows the MVM for the leak $\mathbf{l}_k$ to be executed entirely in the idle time (the offline part) of the processing pipeline, completely removing it from the latency-critical path of the RTC.

A key property of this architecture is that the estimator operates on the accumulated command $\mathbf{m}^*$, which is an internal, exactly known digital quantity, and not on the WFS measurements. Consequently, the regularizer itself is mathematically immune to the PWFS capture range, optical gain variations, and phase wrapping; the method would operate unchanged even if the sensor were entirely blind to the petal modes ($\eta_p = 0$). While the system as a whole remains subject to these nonlinearities---since the sensor determines where the loop locks---the control block is guaranteed to drain the accumulated differential command without triggering instability.

\subsection{Closed-loop stability and robustness of the leaky integrator}
\label{sec:stability}

The closed-loop dynamics are governed by the plant common to all modes---the cascade of the PWFS integration time, the RTC computational latency, and the DM temporal response---which to a good approximation reduces to a pure delay of about three frames in the present configuration.
Throughout this analysis, we assumed that this plant is identical for the continuous and the differential-piston modes; on the ELT, the M4 \citep{2019Msngr.178....3V} response may introduce small mode-dependent differences, but these are negligible for the present discussion. Under this assumption the loop can be studied mode by mode. The continuous modes are controlled by the standard integrator of Eq.~\ref{eq:integrator} and, for the adopted loop gain $g_c$, are asymptotically stable, their closed-loop poles lying strictly inside the unit circle.

The differential-piston subspace is distinct due to how it is sensed. The effective sensitivity $\eta_p$ of the modulated PWFS to the petal modes degrades as the uncorrected residual grows \citep{BertrouCantou2022, 2023A&A...677A.168A}. We note that this loss of sensitivity is driven by the amount of residual wavefront variance rather than by the atmospheric seeing per se. For fragmented ELT pupils, this collapse can be severe \citep{BertrouCantou2022}, though it never vanishes entirely. Writing the closed-loop pole of a generic controlled mode (neglecting the delay, for clarity) as $z \simeq 1 - g_c\,\eta$ (a simplification that captures the dominant pole dynamics even in the presence of loop delays, as formally demonstrated in Appendix \ref{app:delay_analysis}), where $\eta$ is that mode effective sensitivity, the behavior of the petal subspace is seeing-dependent. At good seeing the petal sensitivity $\eta_p$ is sufficient. The petal pole lies inside the unit circle and the standard integrator controls the differential pistons as it does any other mode---which is why the classical control performs excellently in this regime. As the seeing degrades, $\eta_p$ collapses, the pole migrates toward $z=1$, the restoring feedback becomes negligible, and the petal command approaches the regime of a pure integrator,
\begin{equation}
    \label{eq:random_walk}
    p_{k+1} = p_k + d_k,
\end{equation}
driven by an effective disturbance $d_k$ that no longer competes with any genuine error signal. Here $d_k$ aggregates the cross-talk leaking from the well-sensed continuous modes (nonorthogonal to the petal subspace, Sect.~\ref{sec:challenge}), the optical-gain and nonlinear residuals of the sinusoidal PWFS response to phase steps, and the measurement noise projected onto the petal modes. A pole at $z=1$ is marginally---not asymptotically---stable. The loop is not linearly unstable, but the petal command then performs an unbounded random walk whose variance grows without limit. This is the control-theoretic origin of the island effect---a divergence by accumulation rather than by exponential instability---and it explains why the classical integrator remains excellent at good seeing yet fails as the turbulence, and the accompanying loss of petal sensitivity, increase (Sect.~\ref{sec:closed_loop}).

The MMSE soft-limiter acts precisely on this pole. By applying the artificial leak exclusively to the poorly observed petal modes, we prevent their unbounded accumulation. For small values of $g$, the dynamic behavior of the differential piston obeys a straightforward first-order recursion:
\begin{equation}
    \label{eq:leaky_recursion}
    p_{k+1} = (1-g)\,p_k + d_k,
\end{equation}
which is equivalent to replacing the pure integrator by the leaky integrator,
\begin{equation}
    \label{eq:leaky_tf}
    C_\mathrm{petal}(z) = \frac{g_c}{1-(1-g)\,z^{-1}},
\end{equation}
whose controller pole is shifted from $z=1$ to $z=1-g$. Combined with the seeing-dependent sensing feedback, the petal closed-loop pole then sits at $z \simeq (1-g)(1 - g_c\,\eta_p) \approx 1 - g - g_{\mathrm{eff}}$, where $g_{\mathrm{eff}} = g_c\,\eta_p$ is the effective loop gain on the petal subspace. The leak supplies a fixed, sensor-independent restoring term that keeps the pole strictly inside the unit circle even when $\eta_p \to 0$. For $0 < g \ll 1$ the petal mode is therefore asymptotically stable across the whole seeing range, and the random walk turns into a stationary, first-order autoregressive process.

\subsection{Integrator wandering and loop stabilization}
\label{sec:integrator_wandering}
The temporal regularization is fundamentally a trade-off. Introducing a leak on the petal subspace reduces the controller direct-current (DC; i.e., zero-frequency) gain from infinity to $g_c/g$. Including the plant delay of $\tau$ frames, the rejection transfer function (RTF) on the petal subspace becomes
\begin{equation}
    \label{eq:rtf}
    S(z) = \frac{1-(1-g)\,z^{-1}}{1-(1-g)\,z^{-1}+g_{\mathrm{eff}}\, z^{-\tau}},
\end{equation}
which reduces to the pure-integrator form for $g = 0$. At zero frequency ($z = 1$), the pure integrator gives $S = 0$, whereas the leak gives the finite DC rejection factor $S(1) = \rho_{DC} = g/(g + g_{\mathrm{eff}})$. To suppress the control-induced island effect (the random walk described by Eq.~\ref{eq:random_walk}), the leakage gain must be large enough to contain the divergence. For a temporally white disturbance of variance $\sigma_d^2$, the stationary variance of the bounded process is
\begin{equation}
    \label{eq:stationary_var}
    \sigma_p^2 \approx \frac{\sigma_d^2}{2(g + g_{\mathrm{eff}})}.
\end{equation}
Because the injected disturbance $\sigma_d$ grows with the turbulence strength, maintaining an acceptable petal error dictates a performance-driven lower bound on $g$, requiring $g$ to progressively increase as seeing degrades.

\subsection{Continuous modes and cross-talk penalty}
\label{sec:continuous_penalty}
A consequence of this regularization concerns the continuous modes. Because the petal subspace is not perfectly orthogonal to the continuous basis, the correction $g P \hat{\mathbf{p}}$ injects an $\mathcal{O}(g)$ perturbation into the continuous-mode command. 
To first order, the leak matters only at low temporal frequencies. On the petal subspace, the RTF (Eq.~\ref{eq:rtf}) no longer vanishes at zero frequency, as for a pure integrator, but tends to $\rho_{DC}$; above the corner frequency $f_c \approx (g + g_{\mathrm{eff}}) f_s / 2\pi$ the leaky and pure integrators behave identically. The slowly varying, quasi-static part of the turbulent content that the estimator attributes to the petal subspace is therefore only partially corrected, a fraction $\rho_{DC}$ of it remaining in the residual. The resulting variance penalty is
\begin{equation}
    \label{eq:continuous_penalty}
    \Delta\sigma^2 \approx \rho_{\mathrm{DC}}^2\, \sigma^2_{T\to P}(f < f_c), \qquad \frac{\Delta \mathrm{SR}}{\mathrm{SR}} \approx -\left(\frac{2\pi}{\lambda_{\mathrm{sci}}}\right)^2 \Delta\sigma^2,
\end{equation}
where $\sigma^2_{T\to P}$ is the variance of the continuous aberrations (including atmospheric turbulence and unmodeled telescope disturbances such as wind-shake) attributed by the estimator to the petal subspace, and SR is the Strehl ratio. A larger $g$ therefore penalizes the optimal atmospheric correction, an effect we analyze quantitatively via end-to-end simulations in Sect.~\ref{sec:g_optimization}.

\subsection{Quasi-static discontinuities and the retention bound}
\label{sec:lwe_retention}

\begin{figure}[!t]
    \centering
    \includegraphics[width=\columnwidth]{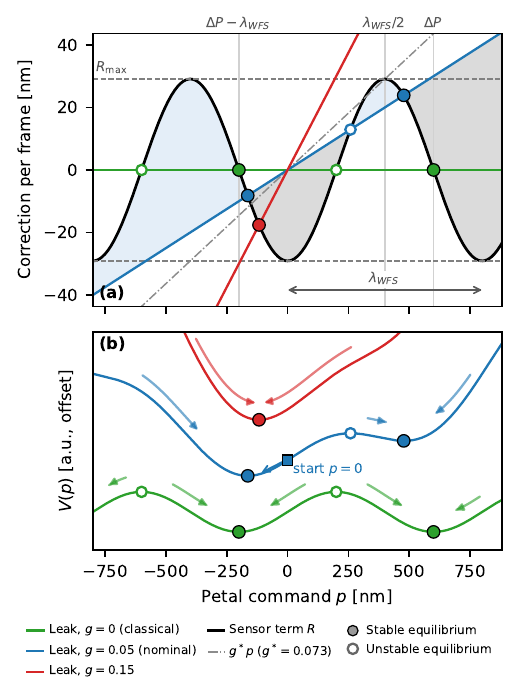}
    \caption{Steady-state equilibria of the petal update (Eq.~\ref{eq:pk1}) for a physical step $\Delta P = 600$~nm ($3\lambda_{WFS}/4$) at $0.8''$ seeing ($g_{\mathrm{eff}} = 0.2285$, $\lambda_{WFS} = 800$~nm). \textbf{(a)} Sensor restoring term $R(\Delta P - p) = R_{\mathrm{max}}\sin[2\pi(\Delta P - p)/\lambda_{WFS}]$, with $R_{\mathrm{max}} = g_{\mathrm{eff}}\lambda_{WFS}/2\pi$ (black; dashed lines at $\pm R_{\mathrm{max}}$), periodic in $\lambda_{WFS}$ (double arrow). The leak is drawn as the drain $g\,p$ subtracted at each frame for $g = 0$ (classical), $g = 0.05$ (nominal), and $g = 0.15$; the equilibria (zero net update) are its crossings with the sensor term, which are stable where the sensor term crosses the drain from above. The vertical gap between the two curves is the net update, and the shading shows its sign for the nominal gain (light blue: $p$ increases; grey: $p$ decreases). The dash-dotted line $g^* p$, with $g^* = g_{\mathrm{eff}}/\pi$, passes through the sensor maximum at $p = \lambda_{WFS}/2$; above $g^*$ no command beyond half a fringe can be sustained. \textbf{(b)} Corresponding washboard potential $V(p)$ (Eq.~\ref{eq:potential}; from bottom to top $g = 0$, $0.05$, $0.15$, curves offset vertically). Stable equilibria are valleys, unstable ones are hills, and arrows show the direction of the update between them. Without a leak all fringe wells are equally deep; the leak acts as a spring anchored at $p = 0$, raising the distant wells until they disappear. At the nominal $g = 0.05$, two stable states coexist ($p \approx 480$~nm and $p \approx -165$~nm); a step applied with the command at $p = 0$ (square) falls in the basin of attraction of the wrapped state, as observed in Fig.~\ref{fig:c1_600nm}. In the limit $\eta_p \to 0$ the washboard flattens: $V(p)$ reduces to a flat line for $g = 0$ (the neutral random walk of Eq.~\ref{eq:random_walk}) and to the single well $\frac{1}{2} g p^2$ for $g > 0$, which keeps the loop bounded but retains no physical step.}
    \label{fig:fixed_point}
\end{figure}

Finally, the fast loop must not blindly filter out true, externally imposed physical phase discontinuities (such as those induced by the low-wind effect). In the presence of a genuine physical step of amplitude $\Delta P$, the response of the modulated PWFS is nonlinear and periodic. The control update for the petal mode at iteration $k$ was modeled as
\begin{equation}
    \label{eq:pk1}
    p_{k+1} = p_k + g_{\mathrm{eff}} \frac{\lambda_{WFS}}{2\pi} \sin\left(\frac{2\pi(\Delta P - p_k)}{\lambda_{WFS}}\right) - g p_k .
\end{equation}
Writing $R(e) = R_{\mathrm{max}}\sin(2\pi e/\lambda_{WFS})$ for the sensor restoring term as a function of the residual $e$, with $R_{\mathrm{max}} = g_{\mathrm{eff}}\lambda_{WFS}/2\pi$, the update reads $p_{k+1} - p_k = R(\Delta P - p_k) - g\,p_k$. The change of the petal command is the sensor correction minus the leak drain.
In writing this isolated modal update, we made two simplifying approximations: we assumed that the static MMSE estimator is unbiased on the petal subspace ($\hat{p} \approx p$), which holds by design in the Gauss-Markov limit, and we neglected the single-frame computational delay of the offline leak calculation.

This update is a gradient descent, $p_{k+1} = p_k - \partial V/\partial p$, on the potential
\begin{equation}
    \label{eq:potential}
    V(p) = \frac{1}{2} g p^2 - g_{\mathrm{eff}}\left(\frac{\lambda_{WFS}}{2\pi}\right)^2 \cos\left[\frac{2\pi(\Delta P - p)}{\lambda_{WFS}}\right],
\end{equation}
and can therefore be read as the overdamped motion of a particle in $V(p)$ (Fig.~\ref{fig:fixed_point}). The two terms of the potential play competing roles. The sinusoidal response of the sensor creates a periodic washboard of period $\lambda_{WFS}$, a sequence of wells (stable equilibria, one per fringe) separated by crests (unstable equilibria), each well being the basin of attraction of one fringe. The leak acts instead as a spring of stiffness $g$ anchored at the zero-petal command, pulling the state toward $p = 0$ with a force $g\,p$ that grows with the distance from the origin.

The spring modifies the washboard in three ways. First, it lifts the degeneracy between fringes. Without a leak, all wells are equally deep and the loop has no preferred fringe, which is the origin of the wandering of the classical integrator, whereas with a leak the wells farther from the origin are progressively raised. Second, it shifts every stable equilibrium toward the origin, which is the droop introduced by the leak ($\approx \rho_{DC}\,\Delta P$ for small steps). Third, it removes equilibria altogether. A well can exist only where the spring force $g|p|$ does not exceed the maximum sensor force $R_{\mathrm{max}}$, so, as $g$ increases each outer well merges with the adjacent crest and both disappear, starting from the most distant ones (in Fig.~\ref{fig:fixed_point}b, two crests for $g = 0$, one for $g = 0.05$, none for $g = 0.15$). Losing the crests means losing the barriers that could hold the state on a distant fringe. When $g \geq g_{\mathrm{eff}}$, that is, when the spring stiffness exceeds the steepest slope of the sensor response, the washboard is fully ironed out and a single, globally stable equilibrium remains for any step, close to, but not exactly at, the zero-petal command.

Applying the existence condition $g|p| \le R_{\mathrm{max}}$ at the edge of the fractional range, $|p| = \lambda_{WFS}/2$, gives the condition for the loop to hold any fractional command:
\begin{equation}
    \label{eq:upper_bound}
    g \lesssim \frac{g_{\mathrm{eff}}}{\pi}.
\end{equation}
The same bound can be read as a retention capacity. The largest command the fast loop can hold in steady state is
\begin{equation}
    \label{eq:pmax}
    p_{\mathrm{max}} = \frac{g_{\mathrm{eff}}\,\lambda_{WFS}}{2\pi g}.
\end{equation}
This bound is not a linear stability limit: no closed-loop pole crosses the unit circle at $g=g_{\mathrm{eff}}/\pi$, and it therefore does not appear in the linear pole analysis of Appendix~\ref{app:delay_analysis}.
Above this threshold, retention degrades gradually: no command beyond half a fringe can be held, and only a fraction of large physical steps is retained.
Together, Eq.~\ref{eq:stationary_var} and Eq.~\ref{eq:upper_bound} show that as seeing degrades, the two boundaries move toward each other, progressively narrowing the operating window of the soft-limiter.
A smaller pyramid modulation radius increases the petal sensitivity $\eta_p$ \citep{2023A&A...677A.168A} and hence $g_{\mathrm{eff}}$. The modulation radius is therefore a design parameter of the soft-limiter, and not only of the continuous-mode sensitivity. Raising $g_{\mathrm{eff}}$ widens the range of $g$ over which full fractional retention holds.

\section{Simulation results}
\label{sec:results}

\subsection{Simulation parameters and baseline configurations}
\label{sec:params}
End-to-end simulations were performed within the SPECULA framework \citep{rossi2026a}, modeling an ELT-like pupil, a six-segment DM, and a $3\lambda/D$ modulated PWFS operating at a sensing wavelength of $\lambda_{WFS} = 800\,\mathrm{nm}$. Performance was evaluated at a science wavelength of $\lambda_{sci} = 2200\,\mathrm{nm}$ (K-band). The system operated at a frame rate of $500\,\mathrm{Hz}$ with a total loop delay of three frames. To represent a realistic operational regime, we assumed a natural guide star of magnitude $R=8$, incorporating both photon and readout noise. This study focuses on the bright-star regime; at fainter magnitudes, the balance between control strategies may shift as the PWFS loses sensitivity to petal modes.

In addition to standard continuous atmospheric turbulence and high-order static aberrations, the simulations incorporated dynamic low-order perturbations originating from the ELT M1 and M2 mirrors using numerical data provided by ESO. The baseline fast loop utilized a modal gain optimization scheme. The continuous modes were governed by a standard integrator with an optimized gain distribution tightly bounded within $g_c \in [0.277, 0.286]$. These values represent the effective loop gains; to achieve this, the digital gains applied in the control law were explicitly scaled by $1/g_{\text{opt}}$, where the modal optical gain $g_{\text{opt}}$ was estimated offline numerically \citep{2023A&A...677A.168A}. This relatively conservative gain regime was dictated by strict stability constraints on a plant model featuring a three-frame delay, designed to guarantee a phase margin of $\approx 50^\circ$ and a gain margin of $\approx 2.2$. We note that this three-frame latency is an approximate assumption for the present study. It simplifies the detailed temporal dynamics of the ELT M4 mirror, and the equivalent delay in the final ANDES system may ultimately be lower. The stabilization result of Appendix~\ref{app:delay_analysis} is independent of this latency, and a lower delay would allow a larger $g_c$ and hence a wider operating window. The present assumption is therefore conservative. The exception is the tip-tilt control, which was handled by a dedicated two-pole IIR filter to reject the M2 wind-shake ($\sim 10\,\mu\mathrm{m}$ RMS). Approximately $33.9\%$ of the reconstructed petal energy falls into this tip-tilt block, meaning the effective petal control is an aggregate of these two temporal paths.

For comparison against standard mitigation strategies, we implemented EAC as the baseline reference. To adequately capture the slow $\sim 0.1\,\mathrm{Hz}$ dynamics of the petal modes and assess fringe lock-in statistics, closed-loop performances were evaluated over extended runs of 30 seconds, utilizing five different atmospheric and noise realizations (with statistics computed over the final 29 seconds to exclude the initial loop-closing transient) to ensure statistically independent assessments.

\subsection{Performance metrics}
\label{sec:metrics}
To unambiguously evaluate the petal control performance, we employed two distinct metrics. The first is the instantaneous Euclidean norm of the five-element differential-piston vector, computed directly from the closed-loop residual phase (taking one segment as the reference):
\begin{equation}
    \|\Delta P(t)\| = \sqrt{ \sum_{i=1}^{5} \Delta P_i(t)^2 }.
\end{equation}
Because this is a norm summed over the modes and not over time (with no mean subtraction), we also report the per-element error $\|\Delta P\|/\sqrt{5}$ to allow a direct comparison with values in the literature. A Euclidean norm does not exhibit $2\pi$ phase wrapping. Therefore, to correctly evaluate the scientific impact in the presence of fringe jumps, we complemented it with the Strehl-based petal error $\sigma_P$ \citep{BertrouCantou2022}:
\begin{equation}
    \sigma_P = \sqrt{-\ln(SR_P)}, \quad \text{with} \quad SR_P = \frac{1}{36}\sum_{i=1}^{6}\sum_{j=1}^{6}\cos\left(\frac{2\pi(a_i - a_j)}{\lambda}\right).
\end{equation}
Evaluating $\sigma_P$ at $\lambda_{WFS} = 800\,\mathrm{nm}$ measures the state of the control loop irrespective of integer fringe jumps, whereas evaluating it at $\lambda_{sci} = 2200\,\mathrm{nm}$ correctly captures the true scientific penalty of any $2\pi$ wrapping.

\begin{table*}[t]
\centering
\caption{Closed-loop quantitative performance comparison of the three control schemes across varying seeing conditions (evaluated over five independent atmospheric and noise realizations).}
\label{tab:performance}
\begin{tabular}{c l c c c c c c c}
\hline\hline
\shortstack{Seeing\\~} & \shortstack{Control\\Scheme} & \shortstack{SR(K)\\~} & \shortstack{$\sigma_P(800)$\\{[rad]}} & \shortstack{$\sigma_P(2200)$\\{[rad]}} & \shortstack{Ratio\\$\frac{\sigma_P(2200)}{\sigma_P(800)}$} & \shortstack{Jump rate\\{[s$^{-1}$]}} & \shortstack{$\max \vert{}\kappa\vert{}$\\~} & \shortstack{$||\Delta P||/\sqrt{5}$\\{[nm]}} \\
\hline
\multirow{3}{*}{$0.6''$}
 & Classical         & 0.910 & 0.157 & 0.057 & \textbf{0.364} & 0.000 & 0 & 33.2 \\
 & EAC               & 0.876 & 0.498 & 0.232 & 0.467 & 0.352$^{(a)}$ & N/A & 144.5 \\
 & MMSE soft-limiter & 0.909 & 0.245 & 0.090 & \textbf{0.367} & 0.000 & \textbf{0} & 54.5 \\
\hline
\multirow{3}{*}{$0.8''$}
 & Classical         & 0.856 & 0.218 & 0.124 & 0.570 & 0.008 & 1 & 72.0 \\
 & EAC               & 0.804 & 0.643 & 0.303 & 0.470 & 0.966$^{(a)}$ & N/A & 187.4 \\
 & MMSE soft-limiter & 0.864 & 0.292 & 0.107 & \textbf{0.368} & 0.000 & \textbf{0} & 64.5 \\
\hline
\multirow{3}{*}{$1.0''$}
 & Classical         & 0.334 & 0.479 & 0.938 & 1.956$^{(b)}$ & 0.741 & 3 & 669.7 \\
 & EAC               & 0.706 & 0.817 & 0.390 & 0.477 & 2.200$^{(a)}$ & N/A & 246.5 \\
 & MMSE soft-limiter & 0.769 & 0.550 & 0.260 & 0.473 & 0.807 & \textbf{1} & 150.5 \\
\hline
\multirow{3}{*}{$1.2''$}
 & Classical         & 0.224 & 0.910 & 1.083 & 1.190$^{(b)}$ & 5.702 & 4 & 794.1 \\
 & EAC               & 0.596 & 0.991 & 0.468 & 0.472 & 5.455$^{(a)}$ & N/A & 291.4 \\
 & MMSE soft-limiter & 0.643 & 0.873 & 0.395 & 0.452 & 4.172 & \textbf{1} & 236.1 \\
\hline
\end{tabular}
\tablefoot{
\tablefoottext{a}{EAC does not actively control the petal modes. Therefore, the jump rate reported for this scheme does not reflect a true fringe jump between locked states, and the integer fringe excursion ($\max \vert{}\kappa\vert{}$) is not applicable (N/A); both metrics merely measure atmospheric wandering across an arbitrary $\pm 400$~nm grid.}
\tablefoottext{b}{Ratios significantly deviating from the physical chromatic scaling ($800/2200 \approx 0.364$) are a direct mathematical signature of unconstrained integer phase wrapping.}
}
\end{table*}

\subsection{Static estimation accuracy}
Before evaluating the closed-loop dynamics, we quantified the baseline precision of the MMSE subspace estimator. This specific test is strictly static. It characterizes the spatial accuracy of the MMSE extraction on independent phase realizations and therefore depends neither on the seeing of a closed loop nor on the temporal leakage gain $g$.

\begin{figure}[h]
    \centering
    \resizebox{\hsize}{!}{\includegraphics{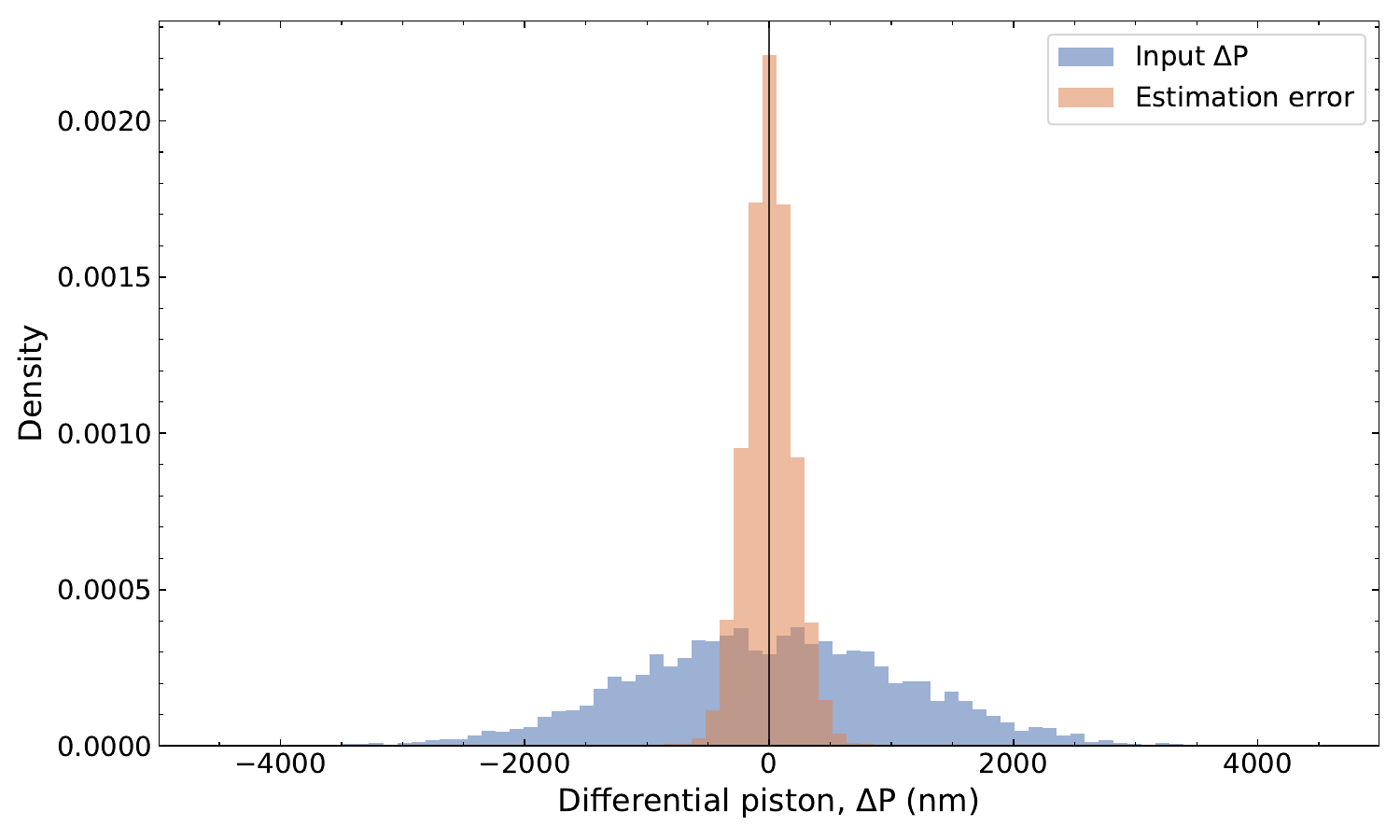}}
    \caption{Static estimation performance. The x-axis shows the differential piston $\Delta P_i = a_i - a_{\text{ref}}$ in nanometers. The plot compares the input $\Delta P$ distribution (blue) and the estimation error after MMSE subspace extraction (orange) over independent phase screens. The input mixture contains turbulence, random petals, and ELT M1/M2 wind-shake, demonstrating the estimator robustness to non-Kolmogorov low-order disturbances.}
    \label{fig:histogram}
\end{figure}

Figure~\ref{fig:histogram} illustrates the statistical distribution of the estimation error. The plotted quantity is the differential piston $\Delta P$ expressed in nanometers. The command vectors from which the petal was extracted were composed of a mix of disturbances: standard atmospheric turbulence, random petal amplitudes, and the ELT M1/M2 wind-shake. The estimator successfully isolates the petal contribution from an input whose low-order content is heavily dominated by a non-Kolmogorov disturbance (the wind-shake) that the von Kármán prior $C_w$ does not inherently model. In our variance breakdown of this static test, the wind-shake accounts for approximately two-thirds of the total error variance. Running the identical test without the wind-shake contribution predictably reduces the residual error per $\Delta P$ element from $218.9\,\mathrm{nm}$ to $126.8\,\mathrm{nm}$ RMS. The extraction degrades gracefully under a misspecified prior, and the loss is essentially fully recovered once the wind-shake statistics are included in $C_w$ (the oracle case, yielding $127.4\,\mathrm{nm}$ RMS).

\begin{figure}[h]
    \centering
    \resizebox{\hsize}{!}{\includegraphics{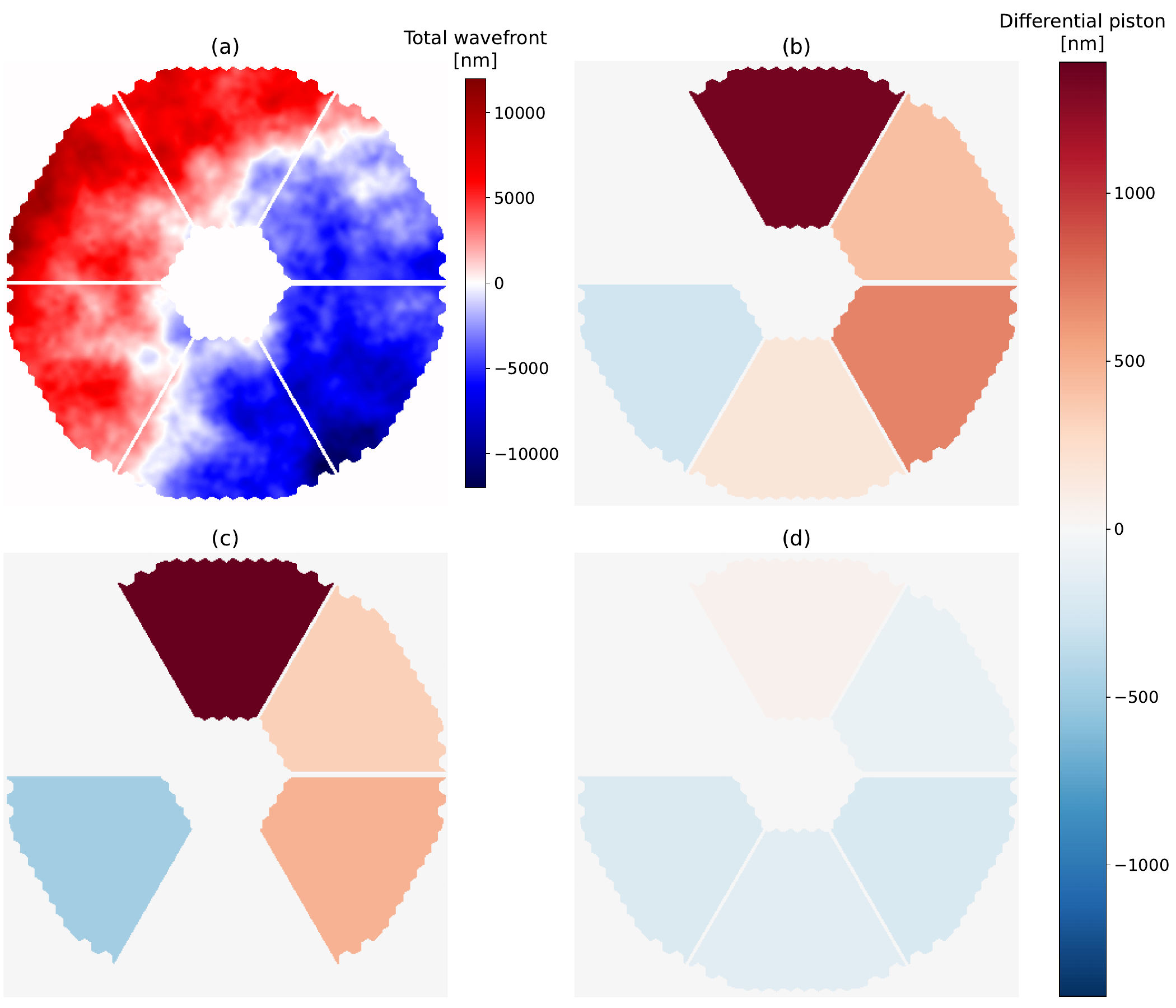}}
    \caption{MMSE subspace extraction for a single static realization. \textbf{(a)} Accumulated total wavefront. \textbf{(b)} Injected ground-truth differential piston. \textbf{(c)} MMSE-estimated differential piston. \textbf{(d)} Estimation error. The RMS error for this specific realization is $157.4\,\mathrm{nm}$.}
    \label{fig:mmse_panels}
\end{figure}

Figure~\ref{fig:mmse_panels} shows a representative single-realization extraction. The narrow residual distribution demonstrates that the linear estimator, utilizing only a single highly optimized MVM, provides a reliable input for the temporal regularizer without requiring nonlinear iterative solvers.

\subsection{Closed-loop temporal stabilization and phase wrapping}
\label{sec:closed_loop}

Building on this static precision, we evaluated the temporal soft-limiter in full closed-loop operations. End-to-end simulations compared three different control schemes under varying seeing conditions (from $0.6$ to $1.2$ arcsec): the classical standard integrator, the EAC baseline, and the proposed MMSE soft-limiter. To ensure a fair comparison, the same optimized modal gain set was derived for each seeing condition following the semi-analytical error budget methodology described in \citet{2019JATIS...5d9001A} and applied identically across all three schemes. We note, however, that the classical control additionally required a modal gain modulation strategy \citep{Pinna2014} during the initial transient to successfully close the loop without diverging. In contrast, both the EAC and the soft-limiter converge without requiring such transient interventions, which is an inherent operational advantage of these regularized architectures. The comprehensive quantitative results of this comparison, including science-band SRs and residual petal errors, are summarized in Table~\ref{tab:performance}.

The absence of phase wrapping at optimal seeing is demonstrated by the chromatic ratio of the Strehl-based petal errors, $\sigma_P(2200)/\sigma_P(800)$. In a purely unwrapped regime, this ratio follows the wavelength scaling ($800/2200 \approx 0.364$). As reported in Table~\ref{tab:performance}, the soft-limiter maintains this physical ratio ($0.367$ at $0.6''$ and $0.368$ at $0.8''$), proving that no integer fringes are accumulated. In contrast, the classical integrator ratio diverges significantly under degrading seeing, signaling severe, unconstrained phase wrapping. Notably, at $0.8''$ the classical integrator already departs from the physical scaling (ratio $0.570$) although its mean fringe-jump rate is only $0.008\,\mathrm{s^{-1}}$ and four out of five seeds remain locked. A single fringe jump in one realization is enough to bias the science-band error. The chromatic ratio is therefore a more sensitive diagnostic of residual wrapping than the jump rate. Furthermore, we explicitly report the maximum integer fringe excursion, $\max \vert{}\kappa\vert{}$, computed from the integer fringe order $\kappa = \mathrm{round}(\Delta P/\lambda_{WFS})$. While the classical control wanders across multiple fringes as seeing degrades, the soft-limiter successfully bounds the residual state. Although it experiences a comparable jump rate to the classical scheme at $1.0''$ (0.807 vs 0.741 s$^{-1}$), the crucial difference lies in the excursion. The soft-limiter confines the differential piston to within a single fringe ($\max \vert{}\kappa\vert{} \le 1$), recovering the adjacent lock basin and avoiding the K-band Strehl collapse.

\begin{figure*}[h]
    \centering
    \includegraphics[width=\textwidth]{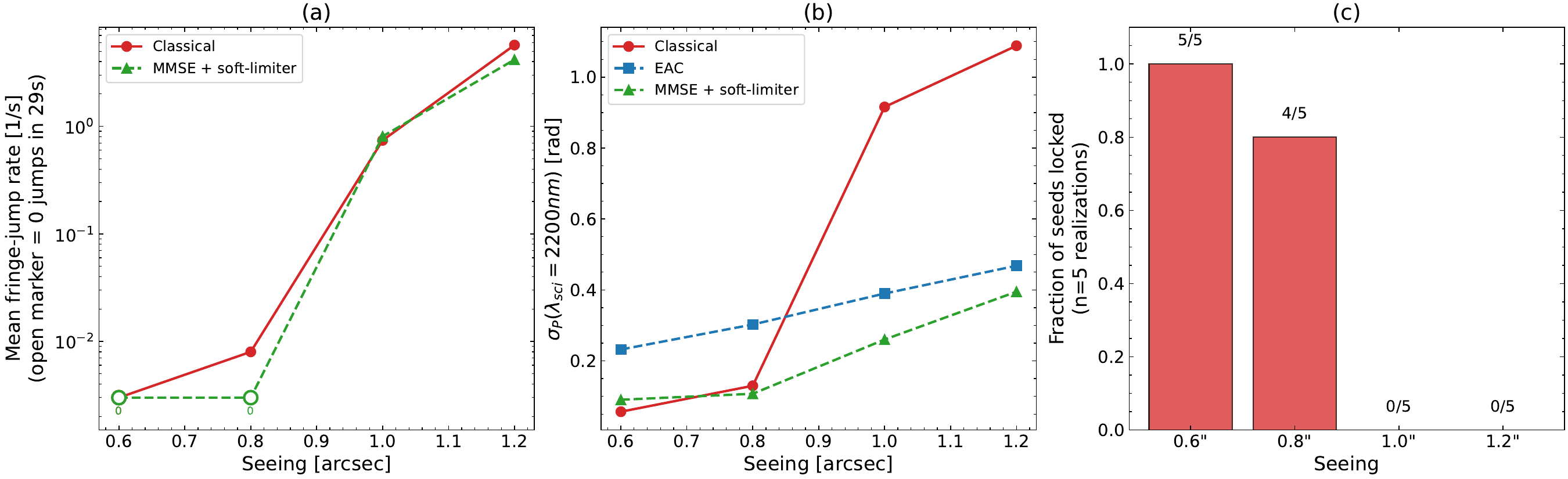}
    \caption{Closed-loop temporal stabilization and phase wrapping analysis. \textbf{(a)} Mean fringe-jump rate over 29 seconds. The EAC does not actively control petal modes, so its transitions reflect atmospheric wandering rather than actual control lock-ins. \textbf{(b)} Science-band petal error $\sigma_P(\lambda_{sci}=2200~\mathrm{nm})$. The MMSE soft-limiter significantly outperforms the classical integrator under severe turbulence without incurring the steady-state performance penalty of the EAC. \textbf{(c)} Fraction of seeds permanently locked in the correct fringe basin for the classical integrator at good seeing (evaluated over five independent atmospheric and noise realizations).}
    \label{fig:fringe_jumps}
\end{figure*}

The results, illustrated in Fig.~\ref{fig:fringe_jumps}, highlight the limitations of standard integrators when facing pupil fragmentation. While the classical control provides excellent stability under optimal seeing conditions (remaining permanently locked in the correct zero-fringe basin, the capture range of the nonlinear sensor response, in five out of five random realizations at $0.6''$), it catastrophically fails as the seeing degrades. At $1.0''$ and $1.2''$, the loss of petal sensitivity $\eta_p$ causes the poorly observed differential pistons to diverge. As shown in Table~\ref{tab:performance} and Fig.~\ref{fig:fringe_jumps}a, at $1.0''$ the MMSE soft-limiter exhibits a jump rate comparable to the classical scheme ($0.807$ vs $0.741\,\mathrm{s}^{-1}$). However, the crucial difference lies in the magnitude of the excursion. The classical integrator wanders freely across several fringes, causing a complete collapse in K-band performance ($\sigma_P(2200) = 0.938\,\mathrm{rad}$), whereas the soft-limiter successfully confines the differential piston to the adjacent lock basin, containing the science-band penalty ($\sigma_P(2200) = 0.260\,\mathrm{rad}$).

Conversely, standard mitigation strategies such as EAC successfully prevent massive petal divergence but inherently degrade the optimal wavefront correction by artificially coupling actuators. Because the EAC forces spatial continuity across the segment gaps, it cannot track independent piston variations, penalizing the SR and elevating the science-band petal error $\sigma_P$ even under optimal seeing conditions (Fig.~\ref{fig:fringe_jumps}b).

At $0.6''$ seeing, the soft-limiter leaves a residual petal error approximately $65\%$ larger than the classical integrator ($54.5\,\mathrm{nm}$ against $33.2\,\mathrm{nm}$ per mode) while maintaining an essentially identical K-band SR ($0.910$ vs $0.909$). The excess is the DC droop of Sect.~\ref{sec:integrator_wandering}. With $g = 0.05$ and $g_{\mathrm{eff}} = 0.348$, the rejection factor is $\rho_{\text{DC}} = 0.126$. The leak deliberately forfeits $\sim 13\%$ of the DC rejection on the petal subspace. This is what bounded behavior at every seeing costs.

As seeing degrades and phase wrapping inevitably occurs (Fig.~\ref{fig:fringe_jumps}a), the soft-limiter successfully confines the state to within one fringe of the truth ($\max \vert{}\kappa\vert{} \le 1$, as reported in Table~\ref{tab:performance}), whereas the classical integrator wanders freely across several fringes. Resolving this bounded residual integer ambiguity remains the intended task of the slow, two-wavelength phasing sensor.

\subsection{Optimization of the leakage gain and operating window}
\label{sec:g_optimization}

As demonstrated in Sect.~\ref{sec:implementation}, the fractional leakage gain $g$ is the only free parameter of the MMSE soft-limiter, and its optimal value is dictated by a strict control-theoretic trade-off. To quantitatively evaluate the theoretical boundary $g \lesssim g_{\mathrm{eff}}/\pi$, the effective loop gain on the petal subspace, $g_{\mathrm{eff}}$, must be accurately known. We determined $g_{\mathrm{eff}}$ by fitting the theoretical RTF of Eq.~\ref{eq:rtf} (with $g = 0$) to the closed-loop temporal power spectral densities of the petal modes. These densities were computed directly from the simulated residual phase (the ground truth) rather than from the WFS telemetry, thus avoiding any circular bias from the optical gain itself. Leaving the delay $D$ as a free parameter, the fit naturally converged to $D = 3.03$ frames, recovering the known plant latency without forcing the model. Fixing $D=3$, we obtained estimates for the effective gains: $g_{\mathrm{eff}} = 0.348$ at $0.6''$, $0.2285$ at $0.8''$, $0.156$ at $1.0''$, and $0.115$ at $1.2''$.

Note that $g_{\mathrm{eff}}(0.6'') = 0.348$ exceeds the nominal integrator gain $g_c \approx 0.28$. As introduced in Sect.~\ref{sec:params}, this does not imply that the petal modes are better sensed than the continuous ones. The two-pole IIR filter controlling the tip-tilt block---which carries $\sim 33.9\%$ of the reconstructed petal energy---has a significantly larger low-frequency gain than the standard integrator. The aggregate effective gain on the petal subspace can therefore exceed the baseline $g_c$. It is precisely for this reason that $g_{\mathrm{eff}}$ must be measured empirically from the temporal transfer functions rather than simply computed as $g_c \eta_p$.

The true petal sensitivity $\eta_p$ fluctuates with the instantaneous residual wavefront variance. Because we extracted $g_{\mathrm{eff}}$ from time-averaged power spectral densities, the resulting values---and consequently the derived upper bounds $g^*$---represent mean operational limits rather than strict instantaneous thresholds. This empirical measurement is independently corroborated by static interaction matrices computed at $3\lambda/D$ modulation~\citep{2023A&A...677A.168A}, which predict a sensitivity ratio $\eta_p(0.6'')/\eta_p(0.8'') = 1.501$, matching our dynamic RTF measurements to within $1.5\%$.

With the bounds established, we performed a parametric scan of $g \in [0.02, 1.0]$ across all seeing conditions to validate Eq.~\ref{eq:stationary_var} and Eq.~\ref{eq:upper_bound}.

\begin{figure*}[h]
    \centering
    \includegraphics[width=\textwidth]{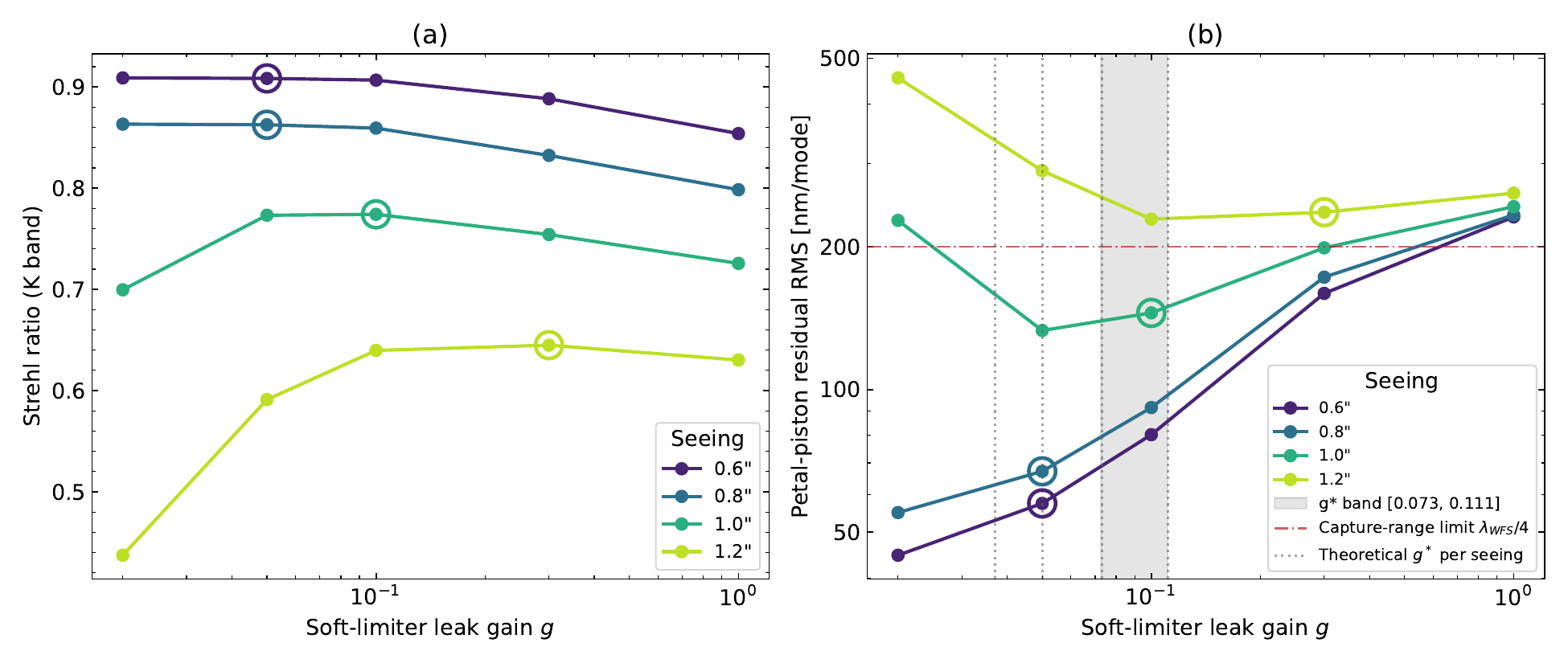}
    \caption{Parametric scan of the soft-limiter leakage gain $g$. \textbf{(a)} K-band SR. \textbf{(b)} Petal residual RMS ($\|\Delta P\|/\sqrt{5}$). Open rings indicate the nominal operating point selected for each seeing condition. The shaded band in (b) represents the theoretical threshold range $g \lesssim g_{\mathrm{eff}}/\pi$ plotted for $0.6''$ and $0.8''$ seeing (where the retention condition is satisfied), while the vertical dotted lines mark the nominal retention threshold, $g^* = g_{\mathrm{eff}}/\pi$, for each corresponding seeing, and the horizontal line marks the capture-range limit $\lambda_{WFS}/4 = 200\,\mathrm{nm}$, beyond which the sinusoidal sensor response is no longer monotonic. Crossing this line while descending in $g$ marks the onset of the divergence described in the text; the rise of all curves as $g \to 1$ has a different origin, and the loop no longer accumulates any petal correction at all. The optimum is strongly asymmetric. Underestimating $g$ under severe turbulence leads to a large loss of Strehl.}
    \label{fig:g_scan}
\end{figure*}

The numerical results in Fig.~\ref{fig:g_scan} confirm the analytical predictions. The need for $g$ to increase with seeing is evident. Under severe turbulence ($1.2''$), the injected atmospheric disturbance $\sigma_d$ is large, requiring a stronger leak to contain the random walk. The optimum of the SR (Fig.~\ref{fig:g_scan}a) is relatively flat but highly asymmetric. This asymmetry has the same nonlinear origin as the upper bound of Sect.~\ref{sec:lwe_retention}. Overestimating $g$ incurs only a minor penalty (1--2\% in SR) due to the $\mathcal{O}(g)$ cross-talk injected into the continuous modes, which grows only linearly with $g$. Underestimating it lets the stationary petal error of Eq.~\ref{eq:stationary_var} grow until it approaches $\lambda_{WFS}/4$, where the sinusoidal sensor response reverses slope. The restoring force collapses, the divergence accelerates, and the loop loses the fringe. For instance, at $1.2''$, setting a too-weak leak ($g=0.02$) fails to stabilize the loop, causing the SR to drop to $0.44$, whereas the nominal value ($g=0.3$) achieves $0.643$. The two edges of the operating window are therefore two faces of the same sensor periodicity --- one deterministic, where the drain imposed by the leak overcomes the maximum restoring signal, the other statistical, where the residual wanders beyond the monotonic range of the sensor.

Guided by this asymmetry, the nominal values for our specific ELT configuration ($g=0.05$ at $0.6''$ and $0.8''$; $g=0.1$ at $1.0''$; $g=0.3$ at $1.2''$) were selected on the higher side of the performance plateau, buying margin against this steep edge, to maximize robustness against unexpected seeing degradations. These stability-driven nominal gains must be compared against the retention bound $g^* = g_{\mathrm{eff}}/\pi$. At $0.6''$ and $0.8''$ seeing, the selected nominal gain ($g=0.05$) safely satisfies the retention threshold ($g^* = 0.111$ and $0.073$, respectively). However, at $1.0''$ and $1.2''$, the values required to contain the random walk ($g=0.1$ and $0.3$) exceed the theoretical thresholds ($g^* \approx 0.050$ and $0.037$). The two boundaries derived in Sect.~\ref{sec:lwe_retention} cross here. From $1.0''$ seeing on, the fast loop retains only a fraction of a physical step ($p_{\mathrm{max}} \approx 199$ and $49\,\mathrm{nm}$ at $1.0''$ and $1.2''$, see Eq.~\ref{eq:pmax}), and the phasing channel must take over the remainder. This crossing has a directly measurable signature. The capture-range condition $\sigma_p \ll \lambda_{WFS}/4 = 200\,\mathrm{nm}$ is met with margin at $0.6''$ and $0.8''$ ($54.5$ and $64.5\,\mathrm{nm}$ per mode, Table~\ref{tab:performance}), is approached at $1.0''$ ($150\,\mathrm{nm}$) and is violated at $1.2''$ ($236\,\mathrm{nm}$). The residual fringe-jump rate reported in Table~\ref{tab:performance} at these seeings is the direct consequence. The soft-limiter bounds the excursion of the differential piston to a single fringe, but above $\approx 1.0''$ it can no longer keep the loop entirely within the monotonic response of the sensor. We also note that as $g \to 1$, the soft-limiter approaches a regime functionally analogous to EAC, in the sense that no petal content is accumulated. However, the mechanism differs. The temporal leak removes the memory, whereas the EAC removes the spatial degree of freedom.
The nominal configuration at $1.2''$ already lies close to this regime. The leak ($g = 0.3$) exceeds $g_{\mathrm{eff}} = 0.115$, so the washboard is fully ironed out (Sect.~\ref{sec:lwe_retention}) and a single equilibrium remains for any step. The fringe jumps measured at this seeing are therefore not locks onto a different fringe, which no longer exist, but transient excursions of the residual driven by the turbulent petal content that the loop no longer retains, as for the EAC; consistently, the two schemes show similar chromatic ratios ($0.452$ and $0.472$) and jump rates at $1.2''$ (Table~\ref{tab:performance}).

\subsection{Consequences for the slow loop (low-wind effect)}
\label{sec:slow_loop}

A critical requirement for the SCAO architecture is that the fast loop must not artificially erase genuine, quasi-static physical steps (such as the low-wind effect), ensuring they remain visible for the slow phasing channel to offload. To evaluate this, we first injected a deterministic square wave perturbation alternating between $0$ and $200\,\mathrm{nm}$ onto two noncontiguous petals during a closed-loop simulation at $0.8''$ seeing.

\begin{figure*}[h]
    \centering
    \includegraphics[width=\textwidth]{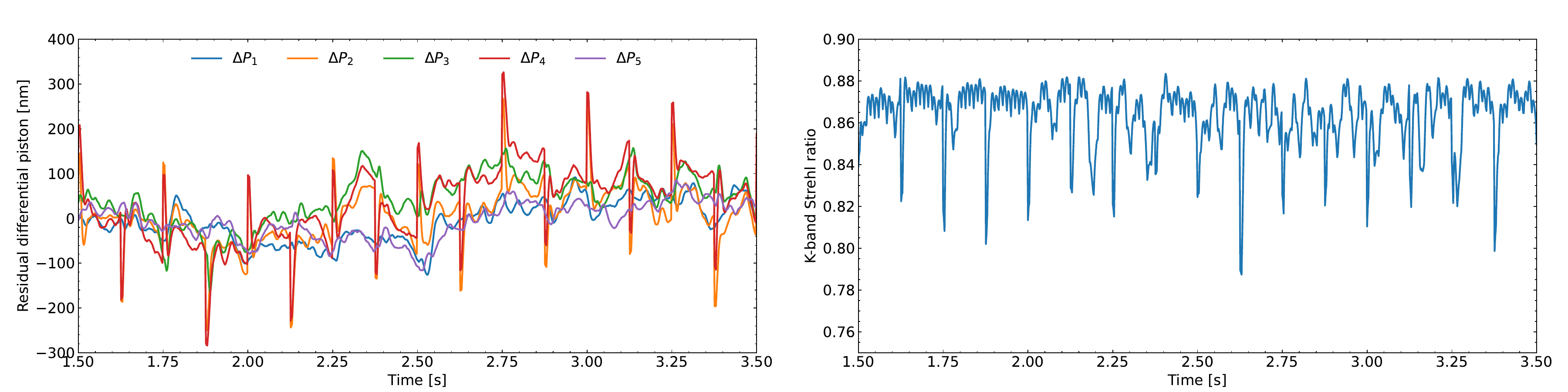}
    \caption{Dynamic response to a 200 nm peak-to-peak square wave injected on two noncontiguous petals at $0.8''$ seeing ($g=0.05$). \textbf{Left:} Temporal history of the five differential petal modes, which faithfully track the injected perturbation. Because the step level is well below the ambiguity threshold, $\lambda_{WFS}/2 = 400$ nm, no phase wrapping occurs. \textbf{Right:} Corresponding K-band SR computed from the total residual phase including the injected disturbance, which remains stable at $\sim 85\%$.}
    \label{fig:c1_200nm}
\end{figure*}

As shown in Fig.~\ref{fig:c1_200nm}, the temporal history of the petal modes faithfully reproduces the injected square wave.
Because the step level ($200\,\mathrm{nm}$) is well below the ambiguity threshold of the sensor ($\lambda_{WFS}/2 = 400\,\mathrm{nm}$), the loop accurately tracks the physical perturbation (with a small $\approx 36\,\mathrm{nm}$ droop) without undergoing phase wrapping.
The soft-limiter successfully drains the control-induced errors while preserving true quasi-static discontinuities, leaving the K-band SR virtually unperturbed.

However, low-wind effect discontinuities can reach significantly larger amplitudes. To verify the robustness boundary derived in Sect.~\ref{sec:lwe_retention}, we repeated the test injecting a step alternating between $0$ and $600\,\mathrm{nm}$ (a $1200\,\mathrm{nm}$ peak-to-peak excursion over a full period cycle).

\begin{figure*}[h]
    \centering
    \includegraphics[width=\textwidth]{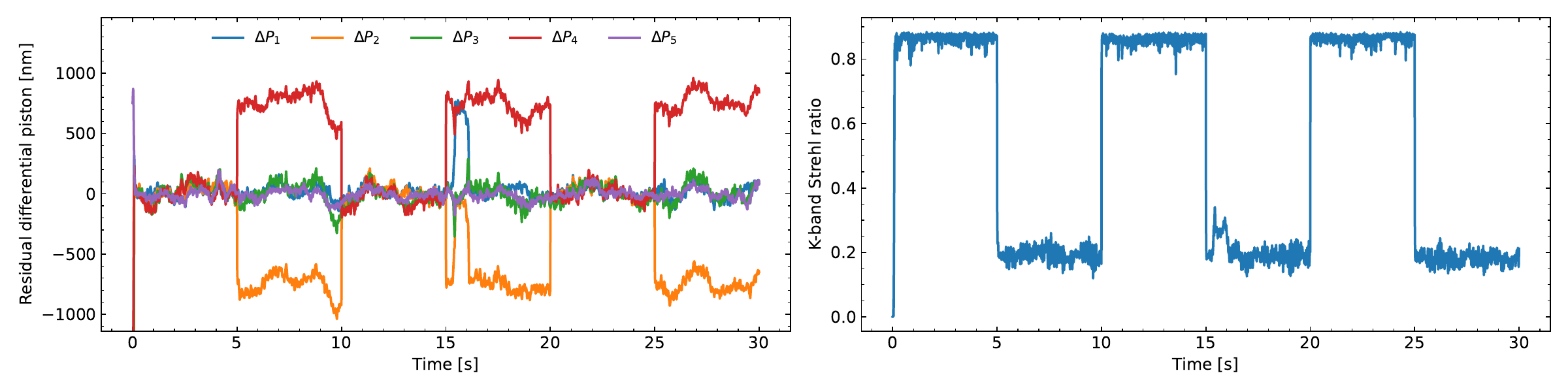}
    \caption{Dynamic response to a severe $0 \to 600\,\mathrm{nm}$ physical step at $0.8''$ seeing ($g=0.05$). \textbf{Left:} Because the step is injected from an initial command, $p=0$, which lies in the basin of attraction of the wrapped well (Fig.~\ref{fig:fixed_point}), the loop settles one fringe away.    The command stably retains the fractional state at $\approx -164\,\mathrm{nm}$ (with a $\approx 36\,\mathrm{nm}$ droop), leaving a residual of $\approx 764\,\mathrm{nm}$ uncorrected. \textbf{Right:} Corresponding K-band SR drops to $\approx 0.2$ for the entire duration of the $600\,\mathrm{nm}$ step due to the uncorrected integer fringe. This demonstrates the fast loop stably holding the fractional state, leaving the integer ambiguity to be resolved by the slow phasing sensor.}
    \label{fig:c1_600nm}
\end{figure*}

Figure~\ref{fig:c1_600nm} demonstrates the capture mechanism under extreme perturbations. Although an unwrapped equilibrium ($p \approx 480\,\mathrm{nm}$) also exists for this $g$ (Fig.~\ref{fig:fixed_point}), the step is applied with the command at $p = 0$, which lies in the basin of the wrapped well; the loop therefore settles one fringe away (as illustrated in Fig.~\ref{fig:fixed_point}b).
The fixed-point model predicts an equilibrium command of $p^* \approx -164\,\mathrm{nm}$ (exhibiting a $\approx 36\,\mathrm{nm}$ droop from the exact $-200\,\mathrm{nm}$ fractional step) and a residual of $\approx 764\,\mathrm{nm}$, which closely matches the residual measured over the steady-state portion of each step (fluctuating between $\sim 748\,\mathrm{nm}$ and $774\,\mathrm{nm}$) in the simulation. No command-space regularizer can do better. The integer fringe is invisible to the sensor. Consequently, the K-band SR drops to $\approx 0.2$ and remains there for the entire duration of the step, recovering fully only when the physical disturbance returns within the lock basin. The quantity the fast loop guarantees is the stability of the fractional state, not the science-band SR; bounding the residual error inside the monotonic capture range ensures that the integer fringe can be safely and unambiguously resolved by the slow phasing sensor on a $\sim 1\,\mathrm{s}$ timescale.

Although the slow channel is not modeled in this work, the washboard picture of Sect.~\ref{sec:lwe_retention} suggests a natural way to couple the two. In the present implementation the spring is anchored at the zero-petal command, so that the basin selected after a large step depends on the history of the loop (Fig.~\ref{fig:fixed_point}b). If the absolute step measured by the phasing channel, $p_{\mathrm{ref}}$, is used instead as the anchor of the leak, i.e., $V(p) = \frac{1}{2} g (p - p_{\mathrm{ref}})^2 - g_{\mathrm{eff}}(\lambda_{WFS}/2\pi)^2 \cos[2\pi(\Delta P - p)/\lambda_{WFS}]$, the deepest well becomes the one of the fringe chosen by the slow channel, and for $g \geq g_{\mathrm{eff}}$ the only one. The slow channel would then select the basin, the leak would enforce it, and the fast loop would track the fractional part within it. Since the leak would then drain only deviations from $p_{\mathrm{ref}}$, the retention limit of Eq.~\ref{eq:pmax} would no longer apply to steps already measured by the phasing channel. Because $p_{\mathrm{ref}}$ varies on the $\sim 1\,\mathrm{s}$ timescale of the phasing sensor, the fast dynamics analyzed here are unaffected. Validating this coupled architecture is left to future work.

\section{Conclusions}
\label{sec:conclusions}

Addressing the differential piston challenge in ELT-class SCAO systems requires navigating mathematical ambiguities. While nonlinear spatial estimators offer high accuracy, the stringent real-time constraints of SCAO strongly favor linear approaches. By shifting the perspective from instantaneous spatial reconstruction to temporal regularization, we have demonstrated that the island effect can be mitigated using a localized leaky integrator based on MMSE von Kármán priors. Our key findings and the theoretical boundaries of this regularization are summarized as follows:

\begin{itemize}
    \item The island effect is a divergence by accumulation. When the petal sensitivity collapses under severe turbulence, the closed-loop pole of the petal subspace migrates to $z=1$ and the poorly observed differential pistons perform a marginally stable random walk. The soft-limiter pulls this pole back inside the unit circle, turning the unbounded divergence into a bounded stationary process.
    \item The behavior depends on the turbulence severity. At good seeing ($\le 0.8''$) the soft-limiter prevents phase wrapping and retains the full fractional range of physical steps, with a small droop. At $1.0''$ occasional fringe jumps occur and steps are only partially retained. At $1.2''$ the nominal leak exceeds $g_{\mathrm{eff}}$, the loop approaches an EAC-like regime, and fringe jumps become frequent. In all cases the excursion remains confined to $\pm 1$ fringe, whereas the classical integrator wanders across several fringes.
    \item The fast loop does not blindly erase physical phase steps. Its dynamics are a gradient descent on a washboard potential, set by the sinusoidal sensor response and confined by the leak, which limits the largest sustainable command to $p_{\mathrm{max}} = g_{\mathrm{eff}}\lambda_{WFS}/(2\pi g)$. For $g \lesssim g_{\mathrm{eff}}/\pi$ the full fractional range up to $\lambda_{WFS}/2$ can be held; above this threshold retention degrades gradually. As seeing degrades, $g$ must increase to contain the random walk while $g_{\mathrm{eff}}$ drops, progressively narrowing the operating window.
    \item At seeing $\le 0.8''$, by keeping the residual inside the monotonic capture range, the soft-limiter provides a stable modulo-$\lambda_{WFS}$ state, free of fringe jumps, on which slow phasing sensors can resolve the integer ambiguity. This is achieved without the baseline continuous-mode penalty of static spatial actuator coupling.
    \item The two channels combine naturally. Anchoring the leak on the step measured by the slow phasing sensor, rather than on the zero-petal command, would let the slow channel select the fringe basin and the fast loop hold the fractional part within it, relaxing the retention limit for steps already measured. This coupled architecture remains to be validated.
    \item The mechanism is not restricted to the petal subspace. Any linear combination of commands that is poorly observed by the WFS can be regularized with its own forgetting factor, computed offline and subtracted at loop rate.
    \item The static subspace estimator can also run passively as a real-time diagnostic, tracking the emergence of petal modes across ELT instruments with minimal computational overhead.
    \item While validated here for SCAO, the approach should carry over to multi-conjugate AO (MCAO) and laser tomography AO (LTAO) systems facing pupil fragmentation.
\end{itemize}

This deterministic approach ensures the necessary stability for secondary, low-frequency phasing sensors to operate effectively, offering a robust baseline applicable not only to the ANDES instrument but potentially to other segmented ELT-class facilities.

\begin{acknowledgements}
      The authors wish to thank the members of the ANDES adaptive optics team for their continuous support and fruitful discussions.
      We are also grateful to Markus Kasper for a stimulating question raised during the AO4ELT8 Conference in Viña del Mar, Chile, which highlighted the operational challenge that inspired this work.
      Finally, the authors acknowledge the use of Gemini (Google) for English language refinement and manuscript formatting.
\end{acknowledgements}

\section*{Data availability}

The end-to-end simulation framework used in this work, SPECULA, is open-source and publicly available. The specific data underlying this article will be shared on reasonable request to the corresponding author (guido.agapito@inaf.it).

\bibliography{report}
\bibliographystyle{aa}

\begin{appendix}
\nolinenumbers
\section{Influence of loop delay on the petal pole dynamics}
\label{app:delay_analysis}

In Sect.~\ref{sec:stability}, the migration of the closed-loop pole was illustrated using a simplified delay-free plant for the sake of clarity. Here, we formally demonstrate that the asymptotic behavior of the dominant pole---which is physically responsible for the control-induced island effect---remains identical when the full temporal delay of the RTC loop is taken into account, using the $z^{-1}$ domain formulation consistent with the main text.

Let $\tau \in \mathbb{N}^+$ represent the total loop latency expressed in units of frames (for instance, $\tau \approx 3$ frames for the SCAO system configuration evaluated in Sect.~\ref{sec:results}). The discrete-time transfer function of the plant representing a single modal command-to-measurement channel can be modeled as
\begin{equation}
    G(z^{-1}) = \eta z^{-\tau},
\end{equation}
where $\eta \in \mathbb{R}$ is the effective optical sensitivity of the PWFS for that specific mode, which degrades under severe residual turbulence or fragmentation. We note that for a sinusoidal step response, $\eta$ is the local slope and can become negative beyond $\lambda_{WFS}/4$; the analysis below is restricted to the locally positive regime, which is the one relevant to the lock basin.

\subsection{Classical pure integrator loop}
When employing a standard pure integrator, the controller transfer function is given by
\begin{equation}
    C_{\text{int}}(z^{-1}) = \frac{g_c}{1 - z^{-1}},
\end{equation}
where $g_c$ is the loop gain. The corresponding closed-loop characteristic equation, defined by $1 + C_{\text{int}}(z^{-1})G(z^{-1}) = 0$, evaluates to
\begin{equation}
    1 + \frac{g_c \eta z^{-\tau}}{1 - z^{-1}} = 0. 
\end{equation}
Multiplying by the denominator yields
\begin{equation}
    \label{eq:char_classical_delay}
    1 - z^{-1} + g_c \eta z^{-\tau} = 0.
\end{equation}
In the critical scenario where the differential piston sensitivity collapses due to severe atmospheric tracking errors ($\eta \to 0$), Eq.~\eqref{eq:char_classical_delay} structurally approaches the limit,
\begin{equation}
    1 - z^{-1} = 0.
\end{equation}
This yields a single unconstrained dominant root located exactly on the unit circle boundary:
\begin{equation}
    z_{\text{dom}} = 1.
\end{equation}
This mathematical limit confirms that the differential piston mode enters a marginally stable random-walk regime characterized by unbounded variance accumulation, regardless of the specific hardware or computational latency $\tau$. The remaining roots of the full characteristic equation are associated with the pure delay and reside safely at the origin ($z = 0$).

\subsection{MMSE soft-limiter loop}
When the proposed temporal regularization is introduced, the controller is transformed into a localized leaky integrator. Its transfer function is expressed as
\begin{equation}
    C_{\text{soft}}(z^{-1}) = \frac{g_c}{1 - (1 - g)z^{-1}},
\end{equation}
where $g$ ($0 < g \ll 1$) is the fractional leakage gain acting as a selective forgetting factor. The modified closed-loop characteristic equation ($1 + C_{\text{soft}}(z^{-1})G(z^{-1}) = 0$) becomes
\begin{equation}
    1 + \frac{g_c \eta z^{-\tau}}{1 - (1 - g)z^{-1}} = 0. 
\end{equation}
Expanding the terms yields
\begin{equation}
    \label{eq:char_soft_delay}
    1 - (1 - g)z^{-1} + g_c \eta z^{-\tau} = 0.
\end{equation}
Applying the same asymptotic loss-of-sensitivity limit ($\eta \to 0$), Eq.~\eqref{eq:char_soft_delay} simplifies into
\begin{equation}
    1 - (1 - g)z^{-1} = 0.
\end{equation}
Solving for $z$, the critical dominant pole is actively bound to
\begin{equation}
    z_{\text{dom}} = 1 - g.
\end{equation}
Because $0 < g < 1$, this dominant root is strictly trapped inside the open unit disk. 

\subsection{Generalization to an arbitrary plant and finite sensitivity}

More generally, let the single-mode plant be $G(z^{-1}) = \eta\,H(z^{-1})$ for an arbitrary stable channel $H$ (loop delay, DM temporal response, etc.). The characteristic equations read $1-z^{-1}+g_c\eta H(z^{-1})=0$ (integrator) and $1-(1-g)z^{-1}+g_c\eta H(z^{-1})=0$ (soft-limiter). For a small but finite sensitivity, the dominant root is, to leading order, $z_\mathrm{dom}\simeq 1-g_c\eta\,H(1)$ and $z_\mathrm{dom}\simeq 1-g-g_c\eta\,H(1)$ respectively, with $H(1)$ the plant DC gain absorbed into the effective sensitivity. These recover the heuristic poles of Sect.~\ref{sec:stability}; in the limit $\eta\to0$ they reduce to $z_\mathrm{dom}=1$ and $z_\mathrm{dom}=1-g$ independently of $H$, and hence of the loop delay. The remaining roots coincide with those of the nominal, well-sensed loop, which is stable by design.

This result proves that the stabilization of the differential-piston mode is decoupled from the loop delay of the AO system. The regularizer applies a localized restoring force that guarantees bounded state variance ($\sigma_p^2 \approx \sigma_d^2/[2(g + g_{\mathrm{eff}})]$, which reduces to $\sigma_d^2/(2g)$ in the limit $\eta \to 0$ considered here; see Eq.~\ref{eq:stationary_var}) even under a temporary total loss of petal sensitivity in the wavefront sensor.

\end{appendix}
\end{document}